\documentclass[conference]{IEEEconf}

\input epsf
\usepackage{titlesec}
\usepackage{balance} 
\usepackage{stfloats}
\usepackage{amsthm}
\usepackage{cite}
\usepackage{amsmath,amssymb,amsfonts}
\usepackage{algorithmic}
\usepackage{graphicx}
\usepackage{textcomp}
\usepackage{xcolor}
\usepackage{booktabs}
\usepackage{multirow}
\usepackage{enumitem}
\usepackage{hyperref}
\usepackage{subcaption}
\usepackage{makecell}

\usepackage{tikz}
\usetikzlibrary{patterns}
\usepackage{pgfplots}
\pgfplotsset{compat=newest}

\theoremstyle{definition}
\newtheorem{definition}{Definition}

\renewcommand\thesection{\arabic{section}} 
\renewcommand\thesubsectiondis{\thesection.\arabic{subsection}}
\renewcommand\thesubsubsectiondis{\thesubsectiondis.\arabic{subsubsection}}

\begin{document}

\title{\textbf{\Large Modelling Android applications through static analysis\\and systematic exploratory testing\\}}

\author{Jordan Doyle$^{1,*}$, Thomas Laurent$^{2}$, and Anthony Ventresque$^{3}$\\
	\normalsize $^{1}$SFI Lero \& School of Computer Science, University College Dublin, Dublin, Ireland\\
	\normalsize $^{2}$JSPS International Research Fellow, National Institute of Informatics, Tokyo, Japan\\
	\normalsize $^{3}$SFI Lero \& School of Computer Science and Statistics, Trinity College Dublin, Dublin, Ireland\\
	\normalsize jordan.doyle@ucdconnect.ie, thomas-laurent@nii.ac.jp, anthony.ventresque@tcd.ie\\
	\normalsize *corresponding author
}

\maketitle

\begin{abstract}
Mobile application development is a fast paced industry with frequent releases. While the development pace increases, so too does the need for automated test generation. Model-based test generation is one of the most common and successful approaches to support this need. Understanding and modelling the application under test is integral to producing comprehensive, dependable and effective tests. Unfortunately, mobile platforms such as Android, introduce a host of difficulties. Static analysis struggles with Android's event-based nature and the growing variety of mechanisms available for developers to implement different features. Additionally, dynamic analysis, implemented by popular random test generators, is slow, inefficient, and limited by a lack of application knowledge.

This paper introduces DroidGraph, a framework to generate a comprehensive control flow model of Android applications using traditional static analysis and efficient systematic exploratory tests. DroidGraph provides a detailed model of an Android application, from low level method statements to high level user interface structures. This model can be used to support automated test generation. We apply DroidGraph to 19 diverse apps and show that our efficient exploratory tests, on average, interact with 18\% more of the app than commonly used random exploration in 345 less interactions. Integrating the dynamic analysis results provided by these tests complements our static analysis, and uncovers on average 51 more components and 49\% more interface callback links in the application code.
\end{abstract}

\IEEEoverridecommandlockouts


\IEEEpeerreviewmaketitle

\section{Introduction}
\label{Section:Inroduction}
Testing is the established technique to ensure stability and reliability in any software-based system. However, with the fast pace commercialisation and increasing marketability of mobile platforms it has become increasingly important to efficiently create strong test suites for complex applications.

Mobile testing remains largely manual, despite the considerable amount of research around test automation~\cite{joorabchi2013real, kochhar2015understanding, linares2017developers, pecorelli2022software, linares2017continuous}. Unit testing, while essential to ensure basic functionality during development, is time consuming and becomes more demanding as the application is developed. Even testing methods designed to reduce a developers workload, such as record and replay, involve consequent manual effort and do not cope well with changing devices and updated code. The majority of research toward Android testing has been devoted to generating user inputs automatically by means of random~\cite{developers2012ui, machiry2013dynodroid}, systematic~\cite{azim2013targeted, moran2017crashscope, amalfitano2012using}, or model-based~\cite{choi2013guided, su2017guided, amalfitano2014mobiguitar} input generation. One of the most well-known automated mobile testing tools is Exerciser Monkey~\cite{developers2012ui}, commonly known as Monkey, which generates pseudo-random streams of user events such as clicks, touches, or gestures in order to test an application. While Monkey's testing method is reliable and requires little maintenance, its random nature introduces significant waste and cannot provide guarantees on the degree of coverage achieved for a certain number of interactions. 

Model-based test generation is the most actively used technique in past research~\cite{kong2018automated}. Although modelling software is not a new area of research, Android introduces new challenges compared to classic sequential execution programs. This makes traditional modelling methods unusable, particularly for modelling callback methods. Mobile applications are highly integrated with the underlying frameworks supported by the mobile operating system, making the execution path and state of an application largely unknown until runtime. Android components such as activity lifecycle methods are called by the Android framework rather than within the application. The framework calls these methods based on the current system state as well as user interactions~\cite{developers2012lifecycle, amalfitano2013testing}. Thus, the majority of models generated in past research are obtained from dynamic analysis.

This paper introduces DroidGraph, a framework to generate a comprehensive control flow model of Android applications using traditional static analysis and efficient systematic exploratory tests. Building on our previous work~\cite{doyle2021improving}, DroidGraph provides a complete model of an Android application, from back-end instruction level code to front-end user interface structures without the need for manual instrumentation. Our previous work using static analysis, manual instrumentation, and control flow model traversal in~\cite{doyle2021improving} suggested that DroidGraph can be used to support automated test generation and provide more efficient tests and better coverage of the application under test (AUT). We now augment our static analysis with efficient systematic exploratory tests on the AUT to enhance the model with runtime data and explore the following questions: 

\begin{enumerate}[label=RQ\arabic*]
    \item How much does the inclusion of dynamic analysis enhance the model of the application compared to pure static analysis?
    \item Can systematically generated exploratory tests reach more of the application than currently used random inputs?
    \item What is the difference in efficiency between systematically and randomly generated exploratory tests?
\end{enumerate}

To study these questions, we apply DroidGraph to 19 diverse apps and show that, on average, our exploratory tests interact with 18\% more of the app than the widely used Exerciser Monkey in 345 less interactions. Integrating the dynamic analysis results provided by these tests complements our static analysis, and uncovers on average 51 more components and 49\% more interface callback links in the application code. The DroidGraph framework as well as the experimental setup and results underlying these conclusions are made available, and can be reproduced using the code and Docker image provided.~\cite{githubrepo}.

The remainder of the paper is structured as follows: Section~\ref{Section:Background} provides the background of Android testing, Section~\ref{Section:Test_Generation} introduces DroidGraph, our control flow model for automated testing of Android apps. Section~\ref{Section:Experiments} details the experiments supporting the exploration of the above mentioned research questions. Section~\ref{Section:Results} describes the results of our experiments and discuss our conclusions from these results. Finally, Section~\ref{Section:Related_Work} provides a summary of related work in the area of model-based Android testing, and Section~\ref{Section:Conclusion} concludes the paper.

\section{Background}
\label{Section:Background}
Testing is the most established technique for maintaining dependable software in any system. The fast pace and ever expanding market of applications available to users makes it especially important for developers to provide a stable and dependable experience. Since Android applications are built with Java, they benefit from being compatible with Java's unit testing frameworks. Unfortunately, this is not the case for all testing related tools. For example, static analysers built for Java will not work with Android because it is compiled into an Android PacKage (APK) using Dalvik bytecode, instead of Java bytecode. An additional conversion from Dalvik to Java bytecode, or some other intermediary representation, is required to make them compatible~\cite{bartel2012dexpler}. 

\begin{figure}[htbp]
\centering\includegraphics[width=\columnwidth]{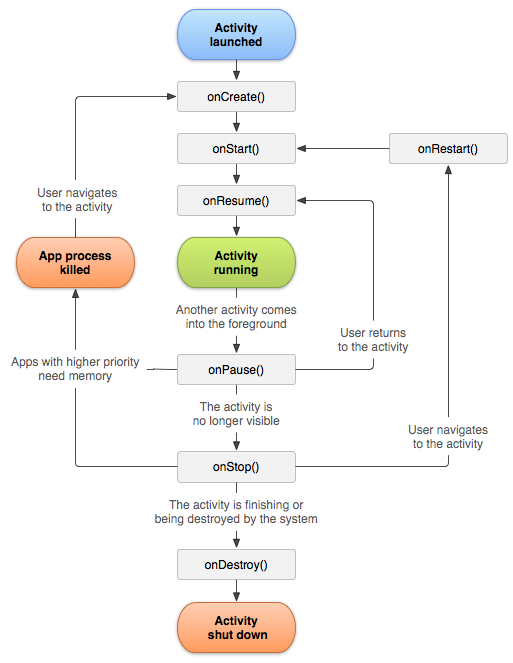}
\caption{A simplified illustration of the activity lifecycle~\cite{developers2012lifecycle}.}
\label{fig:activity_lifecycle}
\end{figure}

Android introduces many challenges to testing, the most significant being the event based nature of the platform and the high integration that individual apps have with the Android framework~\cite{developers2012lifecycle, amalfitano2013testing}. A standard application operates with explicit entry and exit points along with defined paths through the code and interface. Conversely, Android is comprised of standalone units (Activities, Fragments, Services, Broadcasts and Content Receivers) that can be instantiated and referenced in unpredictable ways. This structure lends itself to the interactivity and event based priority of the Android platform. The path taken during execution is largely determined by the Android framework, for example, the order of execution of activity lifecycle methods. Android uses the activity lifecycle to enable developers to ensure that each component functions at every stage of its creation and usage. While the structure of the lifecycle, shown in Figure~\ref{fig:activity_lifecycle}, is clear and concise, its execution is determined by runtime decisions. The activity lifecycle can be entered at any stage; the point at which the application enters is determined by the Android framework~\cite{developers2012lifecycle, amalfitano2013testing}. In a static analysis context, one of the challenges of analysing Android applications is that these lifecycle and listener methods appear disconnected from the rest of the application code. Determining where or when they are called is impossible as it is determined at runtime.

Given the challenges that Android poses, a considerable amount of research has been conducted around test automation, particularly in the area of GUI testing. Many of the challenges associated with Android prevent traditional techniques being applied to Android GUI testing. The event-based priority and GUI-central user experience make interface testing particularly important for mobile applications. Despite considerable research already, mobile testing remains largely manual~\cite{joorabchi2013real, kochhar2015understanding, linares2017developers, pecorelli2022software, linares2017continuous}, making it time consuming and more demanding as the application develops. Frameworks and APIs such as UI Automator~\cite{developers2014uiautomator} as well as many others~\cite{developers2015monkeyrunner, milano2016androidviewclient, developers2016espresso, verma2017mobile, zadgaonkar2013robotium}, allow developers to create test scripts for an applications interface. However, these tools require hours of manual scripting to maintain the test suite as the AUT develops, and scripts are not guaranteed to work on all devices, due to Android device fragmentation~\cite{linares2017continuous}. Record and replay tools~\cite{fazzini2017barista, gomez2013reran, halpern2015mosaic, developer2016espressorecorder} such as Espresso Recorder provide the same features as an automation API but they eliminate the need for scripting. This method provides a quicker way of creating tests but does not solve the core problems; the recordings are still time consuming and reliant on the device used to create them.

The majority of research toward Android testing has been devoted to generating user inputs automatically by means of random~\cite{developers2012ui, machiry2013dynodroid}, systematic~\cite{azim2013targeted, moran2017crashscope, amalfitano2012using}, or model-based~\cite{choi2013guided, su2017guided, amalfitano2014mobiguitar} input generation. Exerciser Monkey~\cite{developers2012ui}, commonly known as Monkey, is the most well known random input generator. It creates pseudo-random streams of interactive events, such as clicks or gestures, and executes them on the AUT. Monkey is reliable and requires little maintenance but it's random strategy introduces waste, and the degree of application coverage cannot be guaranteed. Another well-known random input generator, Dynadroid, applies a Frequency Strategy and Biased-Random Strategy, to increase the efficiency of the selected inputs~\cite{machiry2013dynodroid}. Dynodroid also supports keyboard input and system notifications but these features contribute to its downfall. For instance, in order to provide system events, Dynodroid needs to instrument the Android framework~\cite{machiry2013dynodroid} and in doing so becomes difficult to update and maintain.

Systematic input generation involves dynamically analysing the interface of an application and generating appropriate test inputs as they are found. This method provides effective testing of an application without prior knowledge of the interface structure or the underlying code. Systematic approaches refine random based generators by using sampling strategies and intelligent inputs. Model-based test generation is the most actively used technique in past research~\cite{kong2018automated} and is arguably the most affected by the runtime challenges Android poses as it makes traditional modelling methods unusable, particularly for modelling callback methods. Model-based testing requires a formal model of the application, such as an event flow graph, control flow graph, finite state machine, etc. Once the model has been made, it is used to generate a test suite of device inputs. Due to runtime execution path decisions the majority of models generated in past research are obtained from dynamic analysis.

\section{Control Flow Model}
\label{Section:Test_Generation}
This section first defines the model used by DroidGraph to represent Android applications. It then describes how DroidGraph generates this model from an application using both static and dynamic analysis.

\subsection{Model definition}

\begin{definition}[Extended Control Flow Graph]
An \emph{extended Control Flow Graph} of a program $P$ with a user interface $U$ is a directed graph $G=(V,E)$ where $V=\{v_1,$ $\ldots,$ $v_n\}$, $n\in \mathbb{N}$, is a set of vertices where $v_i$ represents a program point $p_i \in P$ or an interaction $u_i \in U$ and $E=\{e_1, \ldots, e_m\}$, $m\in \mathbb{N}$, is a set of arcs (directed edges), where $e_i=(v_j,v_k)$ with $v_j,v_k \in V$ and $e_i$ representing the flow of control from either, $p_j$ to $p_k$ under some execution of the program $P$ or $u_j$ to $p_k$ after some user interaction.
\end{definition}

DroidGraph's extended control flow graph $G$ is composed of three types of vertices: $V=V_s\cup V_m\cup V_{\mathit{ui}}$, that represent three constituent elements of a mobile application:
\begin{itemize}
\item $V_s$ is the set of statement vertices. While testing, we want to focus on the code written for the application rather than the system or library code. Therefore we exclude such statements from $V_s$. Statements play an important role by revealing application logic, internal method calls, and the detailed control flow of the application.
\item $V_m$ is the set of method vertices. Similarly, it only comprises methods developed for the application. The method vertices represent the entry points to a method and lead to the statement vertices within. These vertices can be further split into three types: Android lifecycle methods, user input callback methods, and standard Java methods.
\item $V_{\mathit{ui}}$ is the set of interface vertices, representing a GUI element a user can interact with and providing input to the application.
\end{itemize}

Not all control flow edges can be included in our graph as they originate from outside of the application code. For example, Activity lifecycle methods are called by the Android framework based on the current state of the system or application, meaning they have no explicit call within the application code. This causes these lifecycle methods to appear as disconnected clusters in the graph. The control flow to and from these clusters is determined at runtime.

\begin{figure}[htbp]
\centering\includegraphics[width=\textwidth, height=0.2\textheight, keepaspectratio]{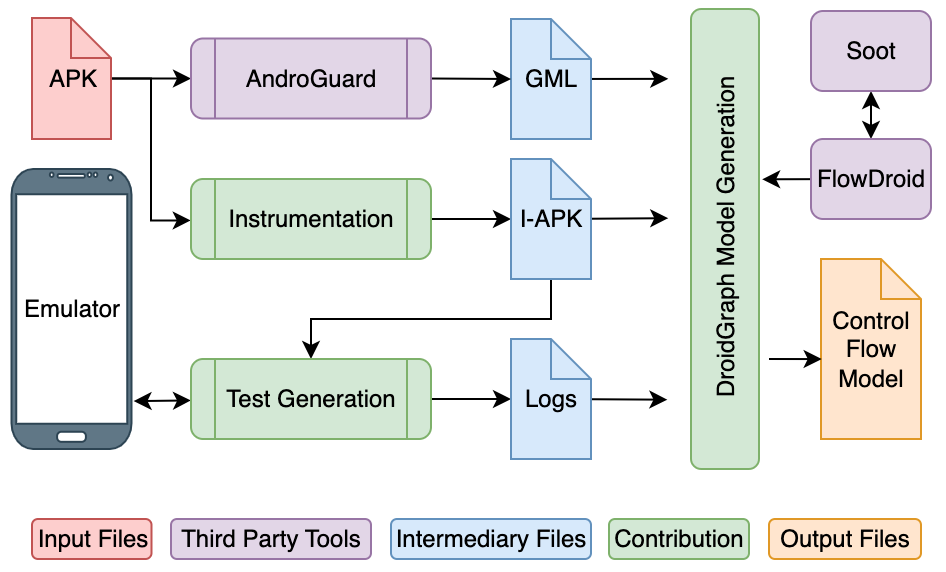}
\caption{DroidGraph component structure.}
\label{fig:overview}
\end{figure}

DroidGraph is composed of several tools and components that work together to create the extended control flow model of an Android application. Figure~\ref{fig:overview} provides an overview of all the processes involved and how they interact. The AUT is provided as input, in the form of an APK file, to Droidgraph's instrumentation module (Section~\ref{Subsection:Instrumentation}) and AndroGuard (Section~\ref{Subsection:AndroGuard}). They produce an instrumented APK file and a GML call graph file, respectively. The instrumented APK is then run with some tests (e.g., exploratory tests generated by DroidGraph, see Section~\ref{Subsection:ExploratoryTests}) which outputs log files containing the actions performed by the tests and any application components found. Using the GML call graph, the instrumented APK and the exploratory test logs, DroidGraph generates the extended control flow model (Section~\ref{Subsection:StaticModelGeneration}). 

\subsection{Instrumentation}\label{Subsection:Instrumentation}

In order to discover links between interface elements and callback methods at runtime, and to calculate both interface- and method-coverage in our experiments, we instrument the AUT using Soot~\cite{lam2011soot}. Soot began as a Java optimisation framework, but by now, researchers and practitioners use Soot to analyse, instrument, optimise and visualise Java and Android applications. Soot's primary functionality is inter-procedural and intra-procedural analysis. It takes as input Java source- and byte-code, or more recently Android byte-code, and transforms it into an intermediary representation known as Jimple---a typed three address code, for easier analysis~\cite{lam2011soot}. Our instrumentation involves adding a log statement to each method of the application (excluding system and library methods), containing the following data: 

\begin{itemize}
\item Log tag: The \texttt{<CONTROL>} tag for any method with a parameter of type \texttt{android.view.View}. The \texttt{<METHOD>} tag for all other methods. Tags are used for filtering logs during and after execution.  
\item Method name: The name of the method is used for 1) linking a callback method to an interface view ID, and 2) calculating method coverage achieved by tests.
\item View ID: Callback methods take a parameter of type \texttt{android.view.View} as input. This is the view object that the user interacts with, and subsequently calls a callback method. If this view object is present, we log the call \texttt{view.getId()}. This enables us to identify which UI element interaction triggered the callback method during runtime. 
\end{itemize}

\subsection{Static Model Generation}\label{Subsection:StaticModelGeneration}

\subsubsection{\textbf{FlowDroid}}

Although Soot has the ability to analyse Android byte-code, it's ability to fully analyse an Android application is limited. It has no knowledge of the Android lifecycle or the Android framework, and therefore cannot analyse or categorise Android-specific callback methods. Unlike Java, Android applications have multiple entry points in the form of callback methods which prevent Soot from creating a call graph. To solve these issues we use FlowDroid~\cite{arzt2014flowdroid}, a fully context-, field-, object- and flow-sensitive taint analysis tool which considers the Android application lifecycle, while also featuring a novel, particularly precise variant of an on-demand alias analysis. It was designed to analyse applications, and alert users of malicious data flows, or as a malware detection tool which could determine if a leak was a policy violation. Despite its main purpose, FlowDroid extends Soot, adding knowledge of the activity lifecycle, and the ability to categorise callback methods; it can thus be used as a powerful static analysis framework.

Our extended control flow graph is composed of three types of vertices, that represent three constituent elements of a mobile application, as defined above. Each type of vertex is collected in a specific way from the initial static analysis:

\begin{itemize}
\item \textbf{Statement} vertices are gathered from FlowDroid \texttt{UnitGraph} objects created for each method in the AUT. Each \texttt{UnitGraph} contains a unit chain detailing all the Jimple statements and their execution order within the method. We also check for inter-procedural calls within each statement of the method.
\item \textbf{Method} vertices are created by retrieving all the methods, for each class, identified by FlowDroid in the AUT. We limit our control flow model to internal components by filtering external library methods such as AndroidX. Before adding methods to the model, we categorise them as activity lifecycle, input listener, or standard Java methods.
\item \textbf{Interface} controls are identified in FlowDroid by parsing the Android XML layout files included in the compiled APK. FlowDroid is capable of identifying controls and their XML attributes when declared in XML layouts but cannot identify controls that have been declared in Java. We attempt to link controls found to their associated listener methods by searching the application code for instances of a \texttt{setListener} method call on a variable containing the resource ID of an interface control. However, due to the number of possible set listener methods, limitations in tracking variable values, and the ambiguity of a developers coding style, preferences and techniques, these links are rarely populated from the static analysis only.
\end{itemize}

\subsubsection{\textbf{AndroGuard}}\label{Subsection:AndroGuard}

Soot is built for analysing Java applications and FlowDroid's main objective is taint analysis, neither is focused on the static analysis of Android applications. This has led to some features becoming outdated and unreliable, for example, the identification of Android fragments. Since FlowDroid was originally released, Android fragments have seen the release of AndroidX, with a newly built API for fragment development. New versions of FlowDroid have yet to add compatibility for fragments built using AndroidX. Consequently, the call graph generated by FlowDroid is consistently missing fragment classes. 

AndroGuard, is a python-based tool used for reverse engineering Android apps~\cite{desnos2012documentation, desnos2011android}, that can produce an acceptable call graph of an Android application. Using AndroGuard, we generate a call graph for the instrumented AUT using the Graph Modelling Language (GML) format. We import the GML graph, filter external methods already classified by AndroGuard, and augment DroidGraph's model with the missing edges between method vertices.

\subsection{Model Enhancement through Dynamic Analysis}\label{Subsection:ExploratoryTests}

There are several control flow components and links missing in the static analysis of Android applications, for example, controls declared within the Java code. While some of these components and links could be retrieved through static analysis, it would require far more development and future maintenance to overcome difficulties such as evolving APIs and developer coding methods. The result would be a system that is not dependable and always behind new trends. Relying on dynamic analysis, through execution of the instrumented application is simple, dependable, and more likely to work with future generations of Android with little to no maintenance.

We explore the AUT using Appium, an open-source project and ecosystem of related software, designed to facilitate UI automation of many app platforms, including Android~\cite{open2012appium, verma2017mobile}. Our exploration mimics an enhanced depth-first search (DFS) on the application GUI. Appium is highly integrated with Android's UI-Automator and provides detailed knowledge of interface elements currently available to interact with, and their locations on the screen. This allows us to avoid interactions that lack corresponding interface element and therefore have no triggered behaviour. Our systematic traversal avoids unnecessary repetitive interactions by tracking interface controls and the number of times they have been used. We also flag interactions that did not lead to further Activities or Fragments (leaf nodes) so that they are not repeated. 

Throughout our exploratory tests we monitor the application logs for instances of our instrumentation, discussed in section~\ref{Subsection:Instrumentation}. When a component that is missing from our model is found it is added. When a log with the tag \texttt{<CONTROL>} is found, we augment our model to include the link between the vertices representing the interface control and listener method found in the log. Droidgraph's exploration of the AUT provides reliable (not subject to randomness) tests to sustain the dynamic analysis phase of model building and complement the static analysis. It does not aim at generating the state of the art, efficient test suites, simply at providing a more reliable alternative to often used random tests such as those generated by Exerciser Monkey.

\section{Experiments}
\label{Section:Experiments}
This section describes the experiments we conducted. First, it describes the research questions we explored. Then it introduces the applications we used as benchmarks. Finally, it describes the protocols and metrics we used to investigate the research questions. The experimental setup and results are made available online~\cite{githubrepo}.

\subsection{Research Questions}\label{sec:rqs}

In order to understand the contribution of DroidGraph we explore the following 3 research questions:
\begin{enumerate}[label=RQ\arabic*]
    \item How much does the inclusion of dynamic analysis enhance the model of the application compared to pure static analysis?

    This research question explores the contribution of the exploratory tests used by DroidGraph to the final model the framework produced. It highlights the elements of the graph that would have been missed if DroidGraph only relied on static analysis. 
    
    \item Can systematically generated exploratory tests reach more of the application than currently used random inputs?

    This research question compares the exploratory power of the inputs generated by DroidGraph and of those generated by Exerciser Monkey. It explores how much of the applications' interface the tests interact with, and how much of their code they execute, reflecting how complete of a model they can lead to.
    
    \item What is the difference in efficiency between systematically and randomly generated exploratory tests?

    This research question studies how quickly the tests generated by DroidGraph and Monkey explore the apps. Even if similar coverage is achieved by both tools, achieving this coverage faster, i.e., using fewer interactions, would lead to faster modelling of the application. 
\end{enumerate}

\subsection{Subject applications}

In order to study the above research questions, we need to apply both DroidGraph and Exerciser Monkey to a diverse set of Android applications. We randomly selected applications from the F-Droid market place~\cite{limited2010fdroid}, which contains over 4000 apps. We select 1 application per category found in F-Droid in order to represent the different types of applications developed in practice. Before choosing the apps we filtered unsuitable apps based on 3 criteria: 

\begin{enumerate}
  \item apps in the games category, as they often include game development frameworks such as unity, which we do not support. 
  \item apps with an Android SDK less than 16 (too old) or greater than 29 (not yet supported by FlowDroid). 
  \item apps that have not been maintained, i.e., not updated within the last 10 years.
\end{enumerate}

We also include 3 apps used in previous research~\cite{doyle2021improving} to show that previous results traversing the control flow model are consistent with our new approach. Table~\ref{tab:subjects} shows the list of apps used in our experiments. All corresponding APK files are made available in~\cite{githubrepo}.

\begin{table}[htbp]
    \centering
    \setlength{\tabcolsep}{2pt}
    \caption{Applications used in the experiment.}
    \label{tab:subjects}
    \begin{tabular}{@{}llr@{}}
         \toprule
         \textbf{Category} & \textbf{App name} & \textbf{Version code} \\
         \midrule
         Connectivity & Webradio & 5 \\
         Development & Git Quick Reference & 7 \\
         Graphics & Pixel Filter & 24 \\
         Internet & Ad-Free & 41 \\ 
         Money & Loyalty Card Keychain & 39 \\
         Multimedia & Camera Roll & 36 \\
         Navigation & GPSTest & 18093 \\ 
         Phone \& SMS & Contact Book & 1 \\ 
         Reading & Drinks & 32 \\
         Science \& Education & Activity Lifecycle & 1 \\
         Science \& Education & Timetable & 17 \\ 
         Security & PIN Mnemonic & 6 \\ 
         Sports \& Health & Wine Cellar & 4 \\
         System & Volume Control & 32 \\
         System & Simple Explorer & 67 \\
         Theming & Battery Live & 13 \\
         Time & MoClock & 4 \\
         Time & Simple Todo & 5 \\
         Writing & Taskkeeper & 6 \\
         \bottomrule
    \end{tabular}
\end{table}

\subsection{Experimental procedure}

In order to investigate RQ1, we first applied DroidGraph to all the applications in Table~\ref{tab:subjects}. During the execution, we recorded which elements in the graph (vertices and edges) were collected from the dynamic analysis results obtained from exploratory tests. We also record the total size of the final model of each application in order to understand the respective contribution of the static analysis and of the dynamic analysis to the model.

In order to investigate RQ2 and RQ3, we also generated exploratory tests for all applications using Exerciser Monkey. As DroidGraph only performs click interactions when exploring the application, we executed Monkey under two settings: limiting it only to the click interaction (Monkey Click), and allowing it to use all interactions it supports (Monkey All). In order to account for the random nature of Monkey, we ran both settings of the tool 10 times for each application. DroidGraph, however, is deterministic in its test generation, and was thus only run once per application. In order to achieve a fair comparison, we ran all tools with the same budget of 500 interactions.

We then measured the overall coverage achieved by each of the generated tests, as well as the coverage achieved after each individual interaction in each test. This was achieved by parsing the logs produced by the instrumentation described in Section~\ref{Subsection:Instrumentation}. Individual interaction coverage is only calculated for the Monkey click only interaction tests due to a limitation in Monkey's output logs. Comparison of the overall coverage achieved for each application by exploratory tests generated by DroidGraph and Monkey serves to answer RQ1, while comparing the rate at which coverage increases during the tests, as well as the number of interactions required to achieve the final coverage answers RQ3. 

All experiments and results can be reproduced using the code and Docker image provided~\cite{githubrepo}.

\section{Results}
\label{Section:Results}
This section describes the results we obtained from our experiments and how they relate to our research questions.  

\subsection{RQ1 How much does the inclusion of dynamic analysis enhance the model of the application?} 

Table~\ref{tab:enhancements} shows the number of vertices and interface-callback edges discovered and added to our extended control flow model thanks to DroidGraph's systematic exploratory tests. On average 52 components are found by the exploratory tests. The majority of these are interface elements declared in Java code rather than Android XML layout files. The largest impact is, on average, 49\% of the missing interface-callback method links, that cannot be retrieved through static analysis, are discovered by our exploratory tests. These results show that while static analysis is effective, Android applications require a combination of static and dynamic analysis and that our efficient systematic exploratory tests are effective in complementing our static analysis in creating a comprehensive control flow model of Android applications.

\begin{table}[tbp]
    \centering
    \setlength{\tabcolsep}{2pt}
    \caption{Dynamic Analysis Model Enhancements.}
    \label{tab:enhancements}
    \begin{tabular}{@{}lrrr@{}}
    \toprule
         \multirow{5}{*}{\textbf{App Name}} & \multicolumn{3}{c}{\textbf{Added elements}}\\
         \cmidrule{2-4}
         & \multirow{2}{*}{\textbf{Vertices}} & \multicolumn{2}{c}{\makecell[b]{\textbf{Interface-Callback}\\\textbf{Edges}}} \\ \cmidrule(l){3-4}
         & \# & \# & \% \\
         \midrule
         Ad-Free & 89 & 89 & 18 \\
         Timetable & 298 & 30 & 26 \\
         Activity Lifecycle & 0 & 24 & 100 \\
         Battery Live & 29 & 32 & 45 \\
         Camera Roll & 18 & 35 & 19 \\
         Contact Book & 7 & 48 & 67 \\
         Drinks & 5 & 7 & 9 \\
         Git Quick Reference & 4 & 7 & 70 \\
         GPSTest & 432 & 23 & 40 \\
         Loyalty Card Keychain & 20 & 38 & 42 \\
         MoClock & 3 & 11 & 91 \\
         PIN Mnemonic & 13 & 41 & 41 \\
         Pixel Filter & 1 & 2 & 13 \\
         Simple Explorer & 12 & 15 & 34 \\
         Simple Todo & 13 & 22 & 49 \\
         Taskkeeper & 8 & 15 & 83 \\
         Volume Control & 5 & 8 & 50 \\
         Webradio & 6 & 7 & 70 \\
         Wine Cellar & 9 & 13 & 62 \\
         \bottomrule
    \end{tabular}
\end{table}

\subsection{RQ2 Can systematically generated exploratory tests discover more of the application than currently used random tests?}

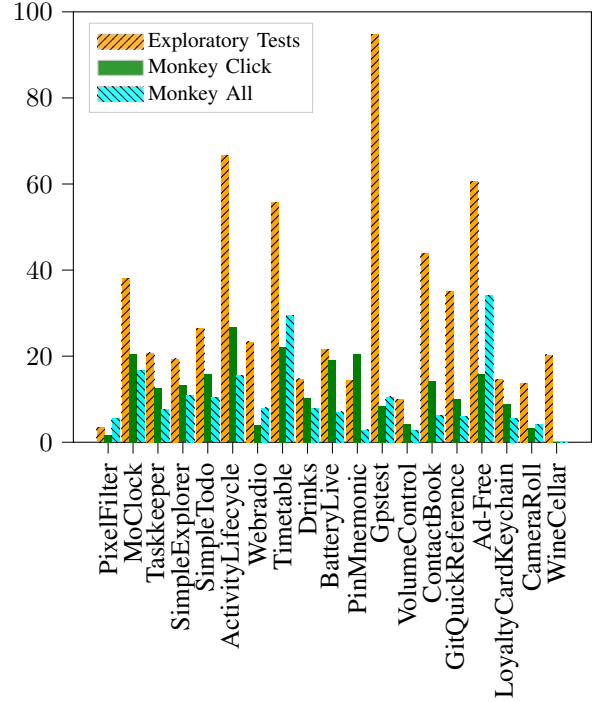
\begin{figure} [tbp]
\begin{tikzpicture}

\definecolor{cyan}{RGB}{0,255,255}
\definecolor{darkgray176}{RGB}{176,176,176}
\definecolor{green}{RGB}{0,128,0}
\definecolor{lightgray204}{RGB}{204,204,204}
\definecolor{orange}{RGB}{255,165,0}

\begin{axis}[
legend cell align={left},
legend style={fill opacity=0.8, draw opacity=1, text opacity=1, draw=lightgray204, at={(0.03,0.87)},
  anchor=west, nodes={scale=0.8, transform shape}},
tick align=outside,
tick pos=left,
x grid style={darkgray176},
xmin=-1.095, xmax=19.695,
xtick style={color=black},
xtick={0.3,1.3,2.3,3.3,4.3,5.3,6.3,7.3,8.3,9.3,10.3,11.3,12.3,13.3,14.3,15.3,16.3,17.3,18.3},
xticklabel style={rotate=90.0},
xticklabels={
  PixelFilter,
  MoClock,
  Taskkeeper,
  SimpleExplorer,
  SimpleTodo,
  ActivityLifecycle,
  Webradio,
  Timetable,
  Drinks,
  BatteryLive,
  PinMnemonic,
  Gpstest,
  VolumeControl,
  ContactBook,
  GitQuickReference,
  Ad-Free,
  LoyaltyCardKeychain,
  CameraRoll,
  WineCellar
},
y grid style={darkgray176},
ymin=0, ymax=100,
ytick style={color=black}
]
\draw[draw=none,fill=orange,postaction={pattern=north east lines}] (axis cs:-0.15,0) rectangle (axis cs:0.15,3.45);
\addlegendimage{ybar,area legend,draw=none,fill=orange,postaction={pattern=north east lines}}
\addlegendentry{Exploratory Tests}

\draw[draw=none,fill=orange,postaction={pattern=north east lines}] (axis cs:0.85,0) rectangle (axis cs:1.15,38.1);
\draw[draw=none,fill=orange,postaction={pattern=north east lines}] (axis cs:1.85,0) rectangle (axis cs:2.15,20.75);
\draw[draw=none,fill=orange,postaction={pattern=north east lines}] (axis cs:2.85,0) rectangle (axis cs:3.15,19.35);
\draw[draw=none,fill=orange,postaction={pattern=north east lines}] (axis cs:3.85,0) rectangle (axis cs:4.15,26.53);
\draw[draw=none,fill=orange,postaction={pattern=north east lines}] (axis cs:4.85,0) rectangle (axis cs:5.15,66.67);
\draw[draw=none,fill=orange,postaction={pattern=north east lines}] (axis cs:5.85,0) rectangle (axis cs:6.15,23.33);
\draw[draw=none,fill=orange,postaction={pattern=north east lines}] (axis cs:6.85,0) rectangle (axis cs:7.15,55.7);
\draw[draw=none,fill=orange,postaction={pattern=north east lines}] (axis cs:7.85,0) rectangle (axis cs:8.15,14.71);
\draw[draw=none,fill=orange,postaction={pattern=north east lines}] (axis cs:8.85,0) rectangle (axis cs:9.15,21.57);
\draw[draw=none,fill=orange,postaction={pattern=north east lines}] (axis cs:9.85,0) rectangle (axis cs:10.15,14.29);
\draw[draw=none,fill=orange,postaction={pattern=north east lines}] (axis cs:10.85,0) rectangle (axis cs:11.15,94.74);
\draw[draw=none,fill=orange,postaction={pattern=north east lines}] (axis cs:11.85,0) rectangle (axis cs:12.15,10);
\draw[draw=none,fill=orange,postaction={pattern=north east lines}] (axis cs:12.85,0) rectangle (axis cs:13.15,43.9);
\draw[draw=none,fill=orange,postaction={pattern=north east lines}] (axis cs:13.85,0) rectangle (axis cs:14.15,35);
\draw[draw=none,fill=orange,postaction={pattern=north east lines}] (axis cs:14.85,0) rectangle (axis cs:15.15,60.67);
\draw[draw=none,fill=orange,postaction={pattern=north east lines}] (axis cs:15.85,0) rectangle (axis cs:16.15,14.68);
\draw[draw=none,fill=orange,postaction={pattern=north east lines}] (axis cs:16.85,0) rectangle (axis cs:17.15,13.73);
\draw[draw=none,fill=orange,postaction={pattern=north east lines}] (axis cs:17.85,0) rectangle (axis cs:18.15,20.31);
\draw[draw=none,fill=green] (axis cs:0.15,0) rectangle (axis cs:0.45,1.72);
\addlegendimage{ybar,area legend,draw=none,fill=green}
\addlegendentry{Monkey Click}

\draw[draw=none,fill=green] (axis cs:1.15,0) rectangle (axis cs:1.45,20.48);
\draw[draw=none,fill=green] (axis cs:2.15,0) rectangle (axis cs:2.45,12.64);
\draw[draw=none,fill=green] (axis cs:3.15,0) rectangle (axis cs:3.45,13.23);
\draw[draw=none,fill=green] (axis cs:4.15,0) rectangle (axis cs:4.45,15.71);
\draw[draw=none,fill=green] (axis cs:5.15,0) rectangle (axis cs:5.45,26.67);
\draw[draw=none,fill=green] (axis cs:6.15,0) rectangle (axis cs:6.45,4);
\draw[draw=none,fill=green] (axis cs:7.15,0) rectangle (axis cs:7.45,22.11);
\draw[draw=none,fill=green] (axis cs:8.15,0) rectangle (axis cs:8.45,10.29);
\draw[draw=none,fill=green] (axis cs:9.15,0) rectangle (axis cs:9.45,19.02);
\draw[draw=none,fill=green] (axis cs:10.15,0) rectangle (axis cs:10.45,20.45);
\draw[draw=none,fill=green] (axis cs:11.15,0) rectangle (axis cs:11.45,8.42);
\draw[draw=none,fill=green] (axis cs:12.15,0) rectangle (axis cs:12.45,4.2);
\draw[draw=none,fill=green] (axis cs:13.15,0) rectangle (axis cs:13.45,14.27);
\draw[draw=none,fill=green] (axis cs:14.15,0) rectangle (axis cs:14.45,10);
\draw[draw=none,fill=green] (axis cs:15.15,0) rectangle (axis cs:15.45,15.73);
\draw[draw=none,fill=green] (axis cs:16.15,0) rectangle (axis cs:16.45,8.72);
\draw[draw=none,fill=green] (axis cs:17.15,0) rectangle (axis cs:17.45,3.14);
\draw[draw=none,fill=green] (axis cs:18.15,0) rectangle (axis cs:18.45,0);
\draw[draw=none,fill=cyan,postaction={pattern=north west lines}] (axis cs:0.45,0) rectangle (axis cs:0.75,5.52);
\addlegendimage{ybar,area legend,draw=none,fill=cyan,postaction={pattern=north west lines}}
\addlegendentry{Monkey All}

\draw[draw=none,fill=cyan,postaction={pattern=north west lines}] (axis cs:1.45,0) rectangle (axis cs:1.75,16.67);
\draw[draw=none,fill=cyan,postaction={pattern=north west lines}] (axis cs:2.45,0) rectangle (axis cs:2.75,7.55);
\draw[draw=none,fill=cyan,postaction={pattern=north west lines}] (axis cs:3.45,0) rectangle (axis cs:3.75,10.97);
\draw[draw=none,fill=cyan,postaction={pattern=north west lines}] (axis cs:4.45,0) rectangle (axis cs:4.75,10.41);
\draw[draw=none,fill=cyan,postaction={pattern=north west lines}] (axis cs:5.45,0) rectangle (axis cs:5.75,15.56);
\draw[draw=none,fill=cyan,postaction={pattern=north west lines}] (axis cs:6.45,0) rectangle (axis cs:6.75,8);
\draw[draw=none,fill=cyan,postaction={pattern=north west lines}] (axis cs:7.45,0) rectangle (axis cs:7.75,29.5);
\draw[draw=none,fill=cyan,postaction={pattern=north west lines}] (axis cs:8.45,0) rectangle (axis cs:8.75,7.94);
\draw[draw=none,fill=cyan,postaction={pattern=north west lines}] (axis cs:9.45,0) rectangle (axis cs:9.75,7.06);
\draw[draw=none,fill=cyan,postaction={pattern=north west lines}] (axis cs:10.45,0) rectangle (axis cs:10.75,2.86);
\draw[draw=none,fill=cyan,postaction={pattern=north west lines}] (axis cs:11.45,0) rectangle (axis cs:11.75,10.53);
\draw[draw=none,fill=cyan,postaction={pattern=north west lines}] (axis cs:12.45,0) rectangle (axis cs:12.75,2.8);
\draw[draw=none,fill=cyan,postaction={pattern=north west lines}] (axis cs:13.45,0) rectangle (axis cs:13.75,6.22);
\draw[draw=none,fill=cyan,postaction={pattern=north west lines}] (axis cs:14.45,0) rectangle (axis cs:14.75,6);
\draw[draw=none,fill=cyan,postaction={pattern=north west lines}] (axis cs:15.45,0) rectangle (axis cs:15.75,34.04);
\draw[draw=none,fill=cyan,postaction={pattern=north west lines}] (axis cs:16.45,0) rectangle (axis cs:16.75,5.6);
\draw[draw=none,fill=cyan,postaction={pattern=north west lines}] (axis cs:17.45,0) rectangle (axis cs:17.75,4.12);
\draw[draw=none,fill=cyan,postaction={pattern=north west lines}] (axis cs:18.45,0) rectangle (axis cs:18.75,0);
\end{axis}

\end{tikzpicture}%
    \caption{Interface coverage (\%) achieved by Exerciser Monkey and by our method}
    \label{fig:interface_bar_chart}
\end{figure}

\begin{figure} [tbp]
\begin{tikzpicture}

\definecolor{cyan}{RGB}{0,255,255}
\definecolor{darkgray176}{RGB}{176,176,176}
\definecolor{green}{RGB}{0,128,0}
\definecolor{lightgray204}{RGB}{204,204,204}
\definecolor{orange}{RGB}{255,165,0}

\begin{axis}[
legend cell align={left},
legend style={fill opacity=0.8, draw opacity=1, text opacity=1, draw=lightgray204, nodes={scale=0.8, transform shape}},
tick align=outside,
tick pos=left,
x grid style={darkgray176},
xmin=-1.095, xmax=19.695,
xtick style={color=black},
xtick={0.3,1.3,2.3,3.3,4.3,5.3,6.3,7.3,8.3,9.3,10.3,11.3,12.3,13.3,14.3,15.3,16.3,17.3,18.3},
xticklabel style={rotate=90.0},
xticklabels={
  PixelFilter,
  MoClock,
  Taskkeeper,
  SimpleExplorer,
  SimpleTodo,
  ActivityLifecycle,
  Webradio,
  Timetable,
  Drinks,
  BatteryLive,
  PinMnemonic,
  Gpstest,
  VolumeControl,
  ContactBook,
  GitQuickReference,
  Ad-Free,
  LoyaltyCardKeychain,
  CameraRoll,
  WineCellar
},
y grid style={darkgray176},
ymin=0, ymax=101.1675,
ytick style={color=black}
]
\draw[draw=none,fill=orange,postaction={pattern=north east lines}] (axis cs:-0.15,0) rectangle (axis cs:0.15,55.32);
\addlegendimage{ybar,area legend,draw=none,fill=orange,postaction={pattern=north east lines}}
\addlegendentry{Exploratory Tests}

\draw[draw=none,fill=orange,postaction={pattern=north east lines}] (axis cs:0.85,0) rectangle (axis cs:1.15,62.5);
\draw[draw=none,fill=orange,postaction={pattern=north east lines}] (axis cs:1.85,0) rectangle (axis cs:2.15,43.15);
\draw[draw=none,fill=orange,postaction={pattern=north east lines}] (axis cs:2.85,0) rectangle (axis cs:3.15,28.26);
\draw[draw=none,fill=orange,postaction={pattern=north east lines}] (axis cs:3.85,0) rectangle (axis cs:4.15,43.05);
\draw[draw=none,fill=orange,postaction={pattern=north east lines}] (axis cs:4.85,0) rectangle (axis cs:5.15,96.35);
\draw[draw=none,fill=orange,postaction={pattern=north east lines}] (axis cs:5.85,0) rectangle (axis cs:6.15,53.79);
\draw[draw=none,fill=orange,postaction={pattern=north east lines}] (axis cs:6.85,0) rectangle (axis cs:7.15,10.55);
\draw[draw=none,fill=orange,postaction={pattern=north east lines}] (axis cs:7.85,0) rectangle (axis cs:8.15,9.77);
\draw[draw=none,fill=orange,postaction={pattern=north east lines}] (axis cs:8.85,0) rectangle (axis cs:9.15,85.21);
\draw[draw=none,fill=orange,postaction={pattern=north east lines}] (axis cs:9.85,0) rectangle (axis cs:10.15,44.54);
\draw[draw=none,fill=orange,postaction={pattern=north east lines}] (axis cs:10.85,0) rectangle (axis cs:11.15,17.86);
\draw[draw=none,fill=orange,postaction={pattern=north east lines}] (axis cs:11.85,0) rectangle (axis cs:12.15,70.34);
\draw[draw=none,fill=orange,postaction={pattern=north east lines}] (axis cs:12.85,0) rectangle (axis cs:13.15,18.74);
\draw[draw=none,fill=orange,postaction={pattern=north east lines}] (axis cs:13.85,0) rectangle (axis cs:14.15,20.05);
\draw[draw=none,fill=orange,postaction={pattern=north east lines}] (axis cs:14.85,0) rectangle (axis cs:15.15,4.99);
\draw[draw=none,fill=orange,postaction={pattern=north east lines}] (axis cs:15.85,0) rectangle (axis cs:16.15,38.42);
\draw[draw=none,fill=orange,postaction={pattern=north east lines}] (axis cs:16.85,0) rectangle (axis cs:17.15,7.48);
\draw[draw=none,fill=orange,postaction={pattern=north east lines}] (axis cs:17.85,0) rectangle (axis cs:18.15,31.2);
\draw[draw=none,fill=green] (axis cs:0.15,0) rectangle (axis cs:0.45,56.06);
\addlegendimage{ybar,area legend,draw=none,fill=green}
\addlegendentry{Monkey Click}

\draw[draw=none,fill=green] (axis cs:1.15,0) rectangle (axis cs:1.45,54);
\draw[draw=none,fill=green] (axis cs:2.15,0) rectangle (axis cs:2.45,35.21);
\draw[draw=none,fill=green] (axis cs:3.15,0) rectangle (axis cs:3.45,22.55);
\draw[draw=none,fill=green] (axis cs:4.15,0) rectangle (axis cs:4.45,36.14);
\draw[draw=none,fill=green] (axis cs:5.15,0) rectangle (axis cs:5.45,55.11);
\draw[draw=none,fill=green] (axis cs:6.15,0) rectangle (axis cs:6.45,48.86);
\draw[draw=none,fill=green] (axis cs:7.15,0) rectangle (axis cs:7.45,5.4);
\draw[draw=none,fill=green] (axis cs:8.15,0) rectangle (axis cs:8.45,9.37);
\draw[draw=none,fill=green] (axis cs:9.15,0) rectangle (axis cs:9.45,80.36);
\draw[draw=none,fill=green] (axis cs:10.15,0) rectangle (axis cs:10.45,71.51);
\draw[draw=none,fill=green] (axis cs:11.15,0) rectangle (axis cs:11.45,5.79);
\draw[draw=none,fill=green] (axis cs:12.15,0) rectangle (axis cs:12.45,40.84);
\draw[draw=none,fill=green] (axis cs:13.15,0) rectangle (axis cs:13.45,7.42);
\draw[draw=none,fill=green] (axis cs:14.15,0) rectangle (axis cs:14.45,15.24);
\draw[draw=none,fill=green] (axis cs:15.15,0) rectangle (axis cs:15.45,3.21);
\draw[draw=none,fill=green] (axis cs:16.15,0) rectangle (axis cs:16.45,18.43);
\draw[draw=none,fill=green] (axis cs:17.15,0) rectangle (axis cs:17.45,3.39);
\draw[draw=none,fill=green] (axis cs:18.15,0) rectangle (axis cs:18.45,20.4);
\draw[draw=none,fill=cyan,postaction={pattern=north west lines}] (axis cs:0.45,0) rectangle (axis cs:0.75,54.04);
\addlegendimage{ybar,area legend,draw=none,fill=cyan,postaction={pattern=north west lines}}
\addlegendentry{Monkey All}

\draw[draw=none,fill=cyan,postaction={pattern=north west lines}] (axis cs:1.45,0) rectangle (axis cs:1.75,40);
\draw[draw=none,fill=cyan,postaction={pattern=north west lines}] (axis cs:2.45,0) rectangle (axis cs:2.75,31.37);
\draw[draw=none,fill=cyan,postaction={pattern=north west lines}] (axis cs:3.45,0) rectangle (axis cs:3.75,23.24);
\draw[draw=none,fill=cyan,postaction={pattern=north west lines}] (axis cs:4.45,0) rectangle (axis cs:4.75,32.15);
\draw[draw=none,fill=cyan,postaction={pattern=north west lines}] (axis cs:5.45,0) rectangle (axis cs:5.75,44.16);
\draw[draw=none,fill=cyan,postaction={pattern=north west lines}] (axis cs:6.45,0) rectangle (axis cs:6.75,48.48);
\draw[draw=none,fill=cyan,postaction={pattern=north west lines}] (axis cs:7.45,0) rectangle (axis cs:7.75,6.27);
\draw[draw=none,fill=cyan,postaction={pattern=north west lines}] (axis cs:8.45,0) rectangle (axis cs:8.75,9.09);
\draw[draw=none,fill=cyan,postaction={pattern=north west lines}] (axis cs:9.45,0) rectangle (axis cs:9.75,62.43);
\draw[draw=none,fill=cyan,postaction={pattern=north west lines}] (axis cs:10.45,0) rectangle (axis cs:10.75,28.07);
\draw[draw=none,fill=cyan,postaction={pattern=north west lines}] (axis cs:11.45,0) rectangle (axis cs:11.75,5.99);
\draw[draw=none,fill=cyan,postaction={pattern=north west lines}] (axis cs:12.45,0) rectangle (axis cs:12.75,46.12);
\draw[draw=none,fill=cyan,postaction={pattern=north west lines}] (axis cs:13.45,0) rectangle (axis cs:13.75,7.47);
\draw[draw=none,fill=cyan,postaction={pattern=north west lines}] (axis cs:14.45,0) rectangle (axis cs:14.75,15.17);
\draw[draw=none,fill=cyan,postaction={pattern=north west lines}] (axis cs:15.45,0) rectangle (axis cs:15.75,3);
\draw[draw=none,fill=cyan,postaction={pattern=north west lines}] (axis cs:16.45,0) rectangle (axis cs:16.75,10.84);
\draw[draw=none,fill=cyan,postaction={pattern=north west lines}] (axis cs:17.45,0) rectangle (axis cs:17.75,4.2);
\draw[draw=none,fill=cyan,postaction={pattern=north west lines}] (axis cs:18.45,0) rectangle (axis cs:18.75,20.52);
\end{axis}

\end{tikzpicture}
    \caption{Method coverage (\%) achieved by Exerciser Monkey and by our method}
    \label{fig:method_bar_chart}
\end{figure}
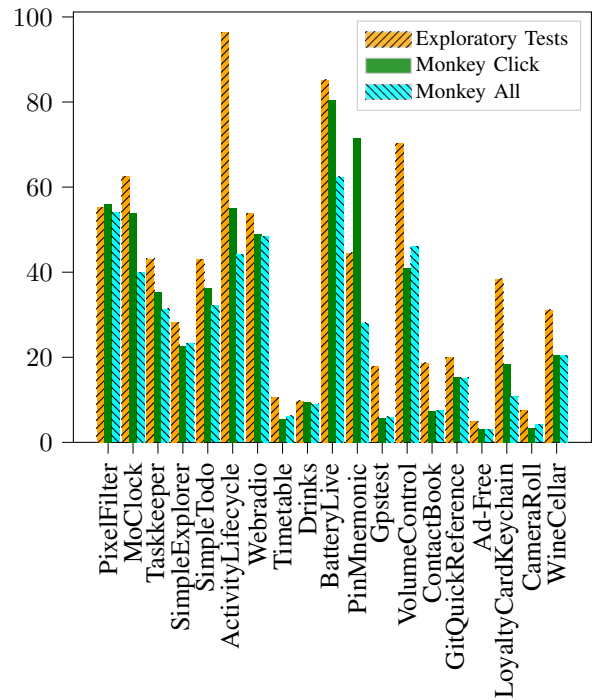

Tables~\ref{tab:interface_coverage} and \ref{tab:method_coverage}, and Figures~\ref{fig:interface_bar_chart} and \ref{fig:method_bar_chart} show a comparison of the coverage achieved by DroidGraph's systematic exploratory tests, as well as the average coverage achieve by both settings of Monkey over the 10 runs. In Tables~\ref{tab:interface_coverage} and \ref{tab:method_coverage}, for each application, the maximum coverage achieved for this app is shown in bold.

DroidGraph's exploratory tests, on average, interact with 18\% more of the AUT than Monkey. While Monkey is comparable in some apps, such as Battery Live, our approach is higher than Monkey's average coverage. Monkey delivered better coverage in only 2 test apps. For example, Pixel Filter achieves the highest coverage in the Monkey all interaction tests. This indicates that Pixel Filter is less click orientated, preventing our exploratory tests and the Monkey click only interaction tests from succeeding.

\begin{table}[tbp]
    \centering
    \caption{Interface coverage (\%) achieved by each approach}
    \label{tab:interface_coverage}
    \setlength{\tabcolsep}{2pt}
    \footnotesize
    \begin{tabular}{@{}lrrrrrrr@{}}
    \toprule
         \multirow{2}{*}{\textbf{App Name}} & \multirow{2}{*}{\makecell{\textbf{Exploratory}\\\textbf{Tests}}} & \multicolumn{3}{c}{\textbf{Monkey Click}} & \multicolumn{3}{c}{\textbf{Monkey All}}\\
         \cmidrule(lr){3-5}\cmidrule(l){6-8}
         && \textbf{Avg} & \textbf{Min} & \textbf{Max} & \textbf{Avg} & \textbf{Min} & \textbf{Max}\\
         \midrule
         Pixel Filter & 3.45 & 1.72 & 0.0 & 3.45 & 5.52 & 0.0 & \textbf{6.9} \\
         MoClock & \textbf{38.1} & 20.48 & 9.52 & \textbf{38.1} & 16.67 & 4.76 & 28.57 \\
         Taskkeeper & \textbf{20.75} & 12.64 & 9.43 & 18.87 & 7.55 & 3.77 & 11.32 \\
         Simple Explorer & \textbf{19.35} & 13.23 & 9.68 & 16.13 & 10.97 & 3.23 & 16.13 \\
         Simple Todo & \textbf{26.53} & 15.71 & 10.2 & 20.41 & 10.41 & 4.08 & 20.41 \\
         Activity Lifecycle & \textbf{66.67} & 26.67 & 2.78 & 44.44 & 15.56 & 8.33 & 22.22 \\
         Webradio & \textbf{23.33} & 4.0 & 3.33 & 6.67 & 8.0 & 3.33 & 10.0 \\
         Timetable & 55.7 & 51.21 & 43.62 & \textbf{59.73} & 29.5 & 0.0 & 48.32 \\
         Drinks & \textbf{14.71} & 10.29 & 0.0 & \textbf{14.71} & 7.94 & 2.94 & \textbf{14.71} \\
         Battery Live & 21.57 & 19.02 & 5.88 & \textbf{33.33} & 7.06 & 1.96 & 11.76 \\
         PIN Mnemonic & 14.29 & 20.45 & 15.18 & \textbf{24.11} & 2.86 & 0.89 & 4.46 \\
         GPSTest & \textbf{94.74} & 8.42 & 5.26 & 10.53 & 10.53 & 10.53 & 10.53 \\
         Volume Control & 10.0 & 4.2 & 4.0 & \textbf{12.0} & 2.8 & 0.0 & 8.0 \\
         Contact Book & \textbf{43.9} & 14.27 & 9.76 & 19.51 & 6.22 & 0.0 & 13.41 \\
         Git Quick Reference & \textbf{35.0} & 10.0 & 10.0 & 10.0 & 6.0 & 5.0 & 10.0 \\
         Ad-Free & \textbf{60.67} & 15.73 & 5.62 & 46.07 & 34.04 & 0.0 & 46.07 \\
         Loyalty Card Keychain & \textbf{14.68} & 8.72 & 9.17 & \textbf{14.68} & 5.6 & 3.67 & 10.09 \\
         Camera Roll & \textbf{13.73} & 3.14 & 1.96 & 4.9 & 4.12 & 1.96 & 7.84 \\
         Wine Cellar & \textbf{20.31} & 0.0 & 0.0 & 0.0 & 0.0 & 0.0 & 0.0 \\
         \bottomrule
    \end{tabular}
\end{table}

\begin{table}[tbp]
    \centering
    \caption{Method coverage (\%) achieved by each approach}
    \label{tab:method_coverage}
    \setlength{\tabcolsep}{2pt}
    \footnotesize
    \begin{tabular}{@{}lrrrrrrr@{}}
    \toprule
        \multirow{2}{*}{\textbf{App Name}} & \multirow{2}{*}{\makecell{\textbf{Exploratory}\\\textbf{Tests}}} & \multicolumn{3}{c}{\textbf{Monkey Click}} & \multicolumn{3}{c}{\textbf{Monkey All}}\\
         \cmidrule(lr){3-5}\cmidrule(l){6-8}
         && \textbf{Avg} & \textbf{Min} & \textbf{Max} & \textbf{Avg} & \textbf{Min} & \textbf{Max}\\
         \midrule
         Pixel Filter & 55.32 & 56.06 & 55.32 & \textbf{58.51} & 54.04 & 52.13 & 56.38 \\
         MoClock & 62.5 & 54.0 & 32.5 & \textbf{70.0} & 40.0 & 30.0 & 55.0 \\
         Taskkeeper & 43.15 & 35.21 & 26.71 & \textbf{44.52} & 31.37 & 30.14 & 32.88 \\
         Simple Explorer & \textbf{28.26} & 22.55 & 18.84 & 23.69 & 23.24 & 20.84 & 27.59 \\
         Simple Todo & \textbf{43.05} & 36.14 & 33.39 & 41.53 & 32.15 & 25.76 & 39.49 \\
         Activity Lifecycle & \textbf{96.35} & 55.11 & 18.98 & 81.75 & 44.16 & 33.58 & 55.47 \\
         Webradio & \textbf{53.79} & 48.86 & 36.36 & 52.27 & 48.48 & 42.42 & 52.27 \\
         Timetable & 10.55 & 5.4 & 8.26 & \textbf{11.5} & 6.27 & 0.0 & 9.96 \\
         Drinks & 9.77 & 9.37 & 8.87 & \textbf{9.83} & 9.09 & 8.36 & 9.75 \\
         Battery Live & \textbf{85.21} & 80.36 & 69.23 & 85.8 & 62.43 & 24.85 & 79.29 \\
         PIN Mnemonic & 44.54 & 71.51 & 65.55 & \textbf{78.15} & 28.07 & 23.53 & 36.13 \\
         GPSTest & \textbf{17.86} & 5.79 & 5.31 & 6.14 & 5.99 & 5.56 & 6.23 \\
         Volume Control & \textbf{70.34} & 40.84 & 49.43 & 63.12 & 46.12 & 41.44 & 49.81 \\
         Contact Book & 18.74 & 7.42 & 3.84 & \textbf{39.17} & 7.47 & 2.92 & 28.26 \\
         Git Quick Reference & \textbf{20.05} & 15.24 & 15.24 & 15.24 & 15.17 & 15.06 & 15.42 \\
         Ad-Free & \textbf{4.99} & 3.21 & 2.59 & 3.83 & 3.0 & 0.0 & 4.31 \\
         Loyalty Card Keychain & \textbf{38.42} & 18.43 & 17.64 & 30.89 & 10.84 & 9.72 & 15.63 \\
         Camera Roll & \textbf{7.48} & 3.39 & 3.33 & 3.68 & 4.2 & 3.29 & 6.31 \\
         Wine Cellar & \textbf{31.2} & 20.4 & 20.4 & 20.4 & 20.4 & 20.4 & 20.4 \\
         \bottomrule
    \end{tabular}
\end{table}

While our exploratory tests are successful in achieving better coverage of the application. The overall coverage achieved is still low in many of the apps. 
There are many reasons for low interface coverage. Interface elements are not always visible until a specific action is executed in the app. For example, a submit button that will only appear after the user enters text into a text field. Some applications require content, such as contact apps, where a portion of the interface is only visible after the user has added a contact. 

\subsection{RQ3 What is the difference in efficiency between systematically and randomly generated exploratory tests?}

\begin{figure} [tbp]
    \input{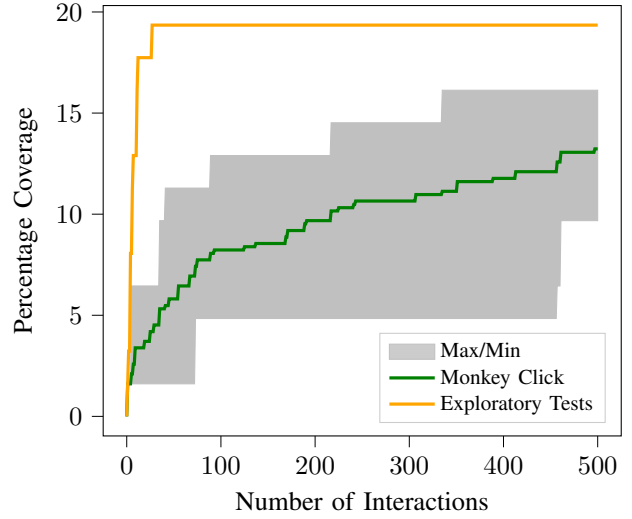}
    \caption{Interface coverage (\%) achieved per interaction by Exerciser Monkey and by our exploratory test on the app Simple Explorer}
    \label{fig:interface_interaction_chart}
\end{figure}

\begin{figure} [tbp]
    \input{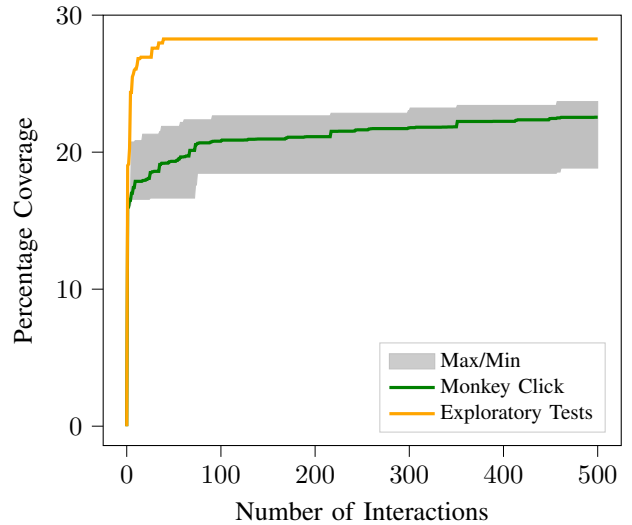}
    \caption{Method coverage (\%) achieved per interaction by Exerciser Monkey and by our exploratory test on the app Simple Explorer}
    \label{fig:method_interaction_chart}
\end{figure}

Figures~\ref{fig:interface_interaction_chart} and \ref{fig:method_interaction_chart} respectively show the interface and method coverage achieved by DroidGraph's tests after each interaction on the application Simple Explorer\footnote{Results for all applications are available in the paper's companion repository~\cite{githubrepo} and left out for space concerns.}, as well as the minimum, maximum, and average coverage achieved by the 10 runs of Monkey (click interactions only) on the app after each interaction. Table~\ref{tab:interaction_coverage} shows the number of interactions required by each approach to achieve its final coverage on each application.

These results show that while Monkey can achieve coverage comparable to DroidGraph's exploratory tests in some cases it is far less efficient in the number of interactions required to achieve it. More interactions required means longer testing times as well as a lot of wasted time performing pointless interactions. DroidGraph's exploratory tests can achieve higher coverage of the AUT in, on average, 345 less interactions. Our exploratory tests execute less interactions than all of Monkey's tests, except for GPSTest and Wine Cellar, which reach their final coverage in only one interaction due to Monkey's inability to cover any of the application.

\begin{table}[tbp]
    \centering
    \caption{Number of interactions required by each approach to achieve maximum coverage of each application.}
    \label{tab:interaction_coverage}
    \begin{tabular}{@{}lrr@{}}
    \toprule
         \textbf{App Name} & \textbf{Exploratory Tests} & \textbf{Monkey Click}\\
         \midrule
         Webradio & 25 & 81 \\
         Git Quick Reference & 71 & 496 \\
         Pixel Filter & 3 & 413 \\
         Ad-Free & 36 & 437 \\ 
         Loyalty Card Keychain & 31 & 497 \\
         Camera Roll & 37 & 499 \\
         GPSTest & 23 & 1 \\ 
         Contact Book & 73 & 493 \\ 
         Drinks & 16 & 488 \\
         Activity Lifecycle & 90 & 492 \\
         Timetable & 5 & 41 \\ 
         PIN Mnemonic & 37 & 493 \\ 
         Wine Cellar & 24 & 1 \\
         Volume Control & 20 & 429 \\
         Simple Explorer & 27 & 461 \\
         Battery Live & 77 & 498 \\
         MoClock & 10 & 489 \\
         Simple Todo & 18 & 496 \\
         Taskkeeper & 109 & 496 \\
         \bottomrule
    \end{tabular}
\end{table}

\section{Threats to Validity}
\label{Section:Threats_Future_Work}
This work's validity is subject to different threats~\cite{Wohlin2012} that this section details, as well as how we addressed them.

\noindent{\bf Construct validity.}
The metrics we use to assess DroidGraph may not be suitable. Although, by construction, the elements added to the model in the dynamic analysis phase are correct (we can not observe a UI element or callback in the logs if it does not exist), the correctness of the model is hard to assess. Similarly, the completeness of the graph can not be assessed due to the complexity of the applications and the lack of ground truth. Instead, we used the coverage of the tests used in the dynamic analysis, a standard metric in testing, as a proxy metric. Indeed, any element covered by the tests will be added to the model, so full coverage would ensure a complete model. Furthermore, by directly counting the number of elements added to the model, we confirm that the dynamic analysis provides a more complete model than the static analysis alone.

\noindent{\bf Conclusion validity.}
Exerciser Monkey relies on randomness to generate tests, which could impact results. To account for this randomness when comparing the systematic exploratory tests used in this work to augment the model through dynamic analysis with the widely used Monkey generated tests, we generated 10 test suites with Exerciser Monkey for each application in our experiments. 

\noindent{\bf Internal validity.}
Results presented in this work could be the consequence of a faulty implementation. To mitigate this threat, we carefully tested the implementation of DroidGraph, and manually verified results where possible.

\noindent{\bf External validity.}
The approach adopted by Droidgraph may not generalise. To mitigate this threat, we assessed it over a set of diverse applications that span multiple domains. We also make the implementation of Droidgraph available, allowing further investigations in the future.

\section{Related work}
\label{Section:Related_Work}
This section outlines previous work in the areas of systematic and model-based Android testing. 

Systematic input generation involves dynamically analysing the interface of an application and generating appropriate test inputs as they are found. This method proves effective in testing an application without prior knowledge of the interface structure or the underlying code. A good example of this technique is AndroidRipper~\cite{amalfitano2012using}, which maintains a state machine model as it systematically traverses the AUT’s UI, called a GUI Tree. This GUI Tree model contains the set of GUI states and state transitions encountered during the ripping process. However, due to a lack of maintenance, AndroidRipper is unusable on newer Android versions. Automatic Android App Explorer (A3E)~\cite{azim2013targeted} implements two forms of exploration, Depth-First Exploration and Targeted Exploration. Depth-First Exploration uses a similar technique to that of AndroidRipper; the framework enters the application at a specified location and explores the application systematically. The aim of this approach is to explore the application in the same manner as a user i.e. by clicking on views to move through the application followed by pressing the back button. Targeted Exploration, conversely, is a directed approach that uses static byte-code analysis to extract a Static Activity Transition Graph, which is then explored systematically while the app runs on the phone. Like AndroidRipper, A3E is poorly maintained and the tool contains functional bugs, for example buttons with labels containing special characters cannot be clicked. Due to its outdated implementation and the need to run the target app under its instrumentation, A3E can cause apps to crash, preventing testing~\cite{wang2018empirical}. Another approach, CrashScope~\cite{moran2017crashscope}, uses static analysis to identify contextual features within activities such as network use. Having identified these features, CrashScope can then test the application in different states, for example with the network on or off. Once static analysis is complete, the tool dynamically extracts the GUI of each screen to identify any clickable, long clickable or text input views, with the intrinsic goal of triggering crashes. While CrashScope shows great potential as an automated testing tool, the main focus of the tool is on the generation of readable crash reports rather than improved effectiveness in detecting bugs. A well-known search-based approach is Sapienz~\cite{mao2016sapienz}, which employs a multi-objective search, combining random fuzzing, systematic and search-based exploration, string seeding and multi-level instrumentation.

Approaches employing systematic input generation show promise, with the aforementioned tools showing varying degrees of improvement in areas such as coverage, input redundancy and bug detection~\cite{wang2018empirical, amalfitano2012using, azim2013targeted, moran2017crashscope, mao2016sapienz}. However, a significant limitation of this technique is its inability to guarantee complete application coverage.
Model-based testing requires a formal model of the application, such as an event flow graph, control flow graph, finite state machine, etc., which is used to generate a test suite of device inputs. Many systematic approaches can be seen as model-based because they create a model while systematically testing the application. Stoat~\cite{su2017guided} and MobiGuitar~\cite{amalfitano2014mobiguitar} model the AUT as a finite state machine (FSM). Stoat uses both static and dynamic analysis, enhanced by a weighted UI exploration strategy, to explore the AUT’s behaviours and construct a stochastic FSM~\cite{su2017guided}. MobiGuitar uses an enhanced version of Android Ripper, dynamically traversing the application in a breadth-first fashion to generate the FSM~\cite{amalfitano2014mobiguitar}. Stoat sets itself apart by using system events within tests. Instead of trying to model system events, it randomly injects various system-level events into its UI tests~\cite{su2017guided}. This simulates the random nature of state changes in a real environment e.g. when notifications are received. Stoat, however, is ineffective with regular gestures (e.g., Pinch Zoom, Move) and specific input data formats~\cite{su2017guided}, while MobiGuitar's reliance on AndroidRipper means it faces the same pitfalls. Another approach to model-based testing is implemented in SwiftHand, which uses machine learning to generate and improve a model of the AUT during test execution. While initially employing a systematic approach, the learned model is used to generate test inputs that visit unexplored states of the AUT. Similar to CrashScope, the main focus of SwiftHand is not on improved fault detection in Android applications. Instead SwiftHand focuses on coverage of the Android AUT and on the reduction of application restarts in automated testing. Application restarts are required by all automatic exploration algorithms, for the exploration of additional states reachable from the initial state, but unlike other learning-based techniques, SwiftHand minimises the number of restarts by attempting to reach unexplored states using only user inputs. SwiftHand’s experimental results show that it can achieve significantly better coverage than traditional random testing in a given time budget, while also reaching peak coverage. However, while SwiftHand excels in reaching coverage goals, the tool cannot handle text input, does not cater for system events and has not been proven capable of handling industry applications.

\section{Conclusions and Future Work}
\label{Section:Conclusion}
The slow and maintenance heavy task of creating test suites coupled with the fast paced mobile industry has increased the need to eliminate manual testing methods as much as possible. Model-based test generation is one of the most common and successful approaches to support automated test generation. Understanding and modelling the test space is integral to producing comprehensive, dependable and effective tests.

This paper introduced DroidGraph, a framework to generate a comprehensive control flow model of Android applications using traditional static analysis and efficient systematic exploratory tests. DroidGraph provides a detailed model of an Android application, from back-end method statements to front-end user interface structures. This model can be used to support automated test generation. We apply DroidGraph to 19 apps spanning 15 categories, and interact with 18\% more of the app using our efficient exploratory tests compared with commonly used random exploration. We also achieve this increase with 345 less interactions. Integrating the dynamic analysis results provided by these tests complements our static analysis, and uncovers on average 51 more components and 49\% more interface callback links in the application code.

Future work will explore the potential of Droidgraph when used with more complex tests that include more interaction types and further explore the application's behaviour in order to evaluate how complete of a model can be created. Furthermore, we also plan to study the use of Droidgraph's application model to generate strong and efficient test suites for Android applications.

\section*{Acknowledgement}
This work was supported, in part, by Science Foundation Ireland grant 13/RC/2094\_P2 and co-funded under the European Regional Development Fund through the Southern \& Eastern Regional Operational Programme to Lero - the Science Foundation Ireland Research Centre for Software (\url{www.lero.ie})

\balance
\bibliographystyle{IEEEtran}
\bibliography{references}

\end{document}